\documentclass[10pt,journal]{IEEEtran}

\usepackage{amsmath,epsfig,amssymb}
\usepackage{bm}
\usepackage{color}
\usepackage{amsthm}
\usepackage[caption=false]{subfig}
\usepackage{cite}
\usepackage[acronym,shortcuts]{glossaries}
\usepackage{lipsum}

\newcommand{\todo}[1]{}
\renewcommand{\todo}[1]{{\color{red} TODO: {#1}}}
\newcommand{\question}[1]{}
\renewcommand{\question}[1]{{\color{red} QUESTION: {#1}}}
\renewcommand\Re{\operatorname{Re}}
\renewcommand\Im{\operatorname{Im}}

\newacronym{GSP}{GSP}{graph signal processing}
\newacronym{GFT}{GFT}{graph Fourier transform}
\newacronym{DGFT}{DGFT}{digraph GFT}
\newacronym{DV}{DV}{directed variation}
\newacronym{IDV}{IDV}{indefinite directed variation}
\newacronym{CDV}{CDV}{complex directed variation}
\newacronym{TV}{TV}{total variation}
\newacronym{IDGFT}{IDGFT}{indefinite DGFT}
\newacronym{CDGFT}{CDGFT}{complex DGFT}
\newacronym{PB}{PB}{protocerebral bridge}
\newacronym{EB}{EB}{ellipsoid body}

\begin{document}

\title{Graph Signal Processing of Indefinite and Complex Graphs using Directed Variation}
\author{\IEEEauthorblockN{Kevin Schultz and Marisel
Villafa\~{n}e-Delgado}~\thanks{This work was supported by NSF award NCS/FO 1835279 and JHU/APL internal
research and development funds}\\
\IEEEauthorblockA{Johns Hopkins University Applied Physics Laboratory\\
11100 Johns Hopkins Road, Laurel, MD, 20723, USA\\
Kevin.Schultz@jhuapl.edu, Marisel.Villafane-Delgado@jhuapl.edu}
}



%

\maketitle
\begin{abstract}

    In the field of graph signal processing (GSP), directed graphs present a particular challenge for the ``standard  approaches'' of GSP to  due to their asymmetric nature.  The presence of negative- or complex-weight directed edges, a graphical structure used in fields such as neuroscience, critical infrastructure, and robot coordination, further complicates the issue.  Recent results generalized the total variation of a graph signal to that of directed variation as a motivating principle for developing a graphical Fourier transform (GFT).  Here, we extend these techniques to concepts of signal variation appropriate for indefinite and complex-valued graphs and use them to define a GFT for these classes of graph.  Simulation results on random graphs are presented, as well as a case study of a portion of the fruit fly connectome.

\end{abstract}

\section{Introduction}


Given a network represented by a graph, where vertices correspond to components and edges represent some relationship, a \textit{graph signal} is a function or time series defined on the graph vertices. \Ac{GSP} is a developing technique for understanding dynamics  evolving on discrete network structures
\cite{shuman2013emerging,sandryhaila2013discrete}.
Recent applications include the analysis of interconnections between components in the brain \cite{raj2019spectral} and the interpretability of deep networks in machine learning \cite{anirudh2017margin}. \ac{GSP} analysis has largely focused on extending signal processing concepts for processing graph signals, which are no longer defined on regular (Euclidean) domains. Most techniques trace their roots to the fields of algebraic signal processing \cite{puschel2008algebraic} or spectral graph theory \cite{chung1997spectral}.

The \ac{GFT} is a fundamental operator of graph signals, where now the basis functions and frequencies are defined from the eigenvectors and eigenvalues of a graph matrix (e.g., graph Laplacian), respectively. In fact, for so-called \textit{regular} graphs such as lines, rings, and grids, this approach is identical to the standard discrete Fourier transform with appropriate dimension and boundary conditions.  The basic definition of the \ac{GFT} \cite{shuman2013emerging} assumes undirected graphs with positive edges. Recent work developed a practical \ac{GFT} for directed graphs \cite{shafipour2018directed}, named \ac{DGFT}. In this approach, a \ac{DGFT} matrix is obtained from optimization methods that aim to obtain graph frequencies as uniformly distributed as possible. They introduce a two-step optimization procedure, which first aims to find the maximum \ac{DV}, a generalization from the \ac{TV}. The second step seeks to minimize a spectral dispersion function, finding basis elements whose distribution of \ac{DV} is smooth and fall within the achievable frequency range in order to obtain a spread \ac{DGFT} basis. 

For many applications of interest, the underlying network structure is directed or the edge weights can be negative or complex, and then the adjacency matrix is no longer positive and symmetric.  Under these conditions, the eigenvectors of the Laplacian $\bm{L}$
no longer form a valid \ac{GFT} basis. The work in  \cite{shafipour2018directed} only considers positive weights, and this is not compatible with the network structure in many applications, such as biological neural networks which are best modeled 1) as directed graphs through pre-synaptic to post-synaptic connections and 2) indefinite (i.e., positive and negative) weighted graphs due to the presence of excitatory and inihibitory neurons.  Furthermore, due to Dale's law \cite{eccles1976electrical}, each vertex should have only either positive or negative weights leaving it (with few exceptions), imposing additional structure on the graph.

In this paper we extend the work in  \cite{shafipour2018directed} for directed graphs with positive, negative, or complex weights and/or graph signals. We introduce novel concepts of \ac{IDV} and \ac{CDV}. Furthermore, we introduce appropriate modifications for the greedy and feasible optimization strategies introduced in  \cite{shafipour2018directed} based on \ac{IDV} and \ac{CDV}. 

In the following sections, we first review related work on \ac{GSP} on directed graphs, in particular focusing on recent results from \cite{shafipour2018directed}.  Next, we show how these concepts extend to \ac{IDGFT} and \ac{CDGFT} to perform \ac{GSP} on indefinite or complex-weighted graphs and address the necessary modifications to the algorithms in \cite{shafipour2018directed}.  We exercise these new techniques on a set of randomly generated graphs, and finally perform a case-study of a particular indefinite directed graph that models dynamics in a portion of the fruit fly connectome.

\section{Background}

In the most basic framework of \ac{GSP}, we are given a positive-weighted,
undirected graph $\mathcal{G}=(\mathcal{V},\bm{A})$, where $\mathcal{V}$ is the
set of vertices with $|\mathcal{V}|=N$ and $\bm{A}\in\mathbb{R}^{N\times N}$ is the adjacency matrix of $\mathcal{G}$ with $A_{ij}\geq0$.  Additionally, a real valued function $\bm{x}:\mathcal{V}\to\mathbb{R}^N$ is defined on the vertices of $\mathcal{G}$. 
A common approach  \cite{shuman2013emerging} is to project $\bm{x}$ onto the eigenvectors of the graph Laplacian $\bm{L}\triangleq \bm{D}-\bm{A}$, where $\bm{D}$ is the diagonal degree matrix, i.e., $D_{ii}=\sum_j A_{ji}$.

When $\bm{A}$ is nonnegative-symmetric, the eigenvalues $\lambda_i$ of
$\bm{L}$ satisfy $0=\lambda_1<\lambda_2\leq\cdots\leq\lambda_N$ and the
corresponding eigenvectors $v_i$ are linearly independent.  This leads to the
analogy that $\lambda_i$ corresponds to frequency and $v_i$ corresponds to the Fourier harmonics.  Letting
$V=[v_1,\dots,v_N]$ defines a \ac{GFT} $\tilde{\bm{x}}=V^\top \bm{x}$,
that reduces to the classical discrete time Fourier transform when
$\mathcal{G}$ is an unweighted ring graph. This insight is the intuition for further analogues from classical signal processing.

When the graph $\mathcal{G}$ is directed, this basic formulation using $\bm{L}$ breaks down. First, there is no uniform definition of a
Laplacian for a directed graph.  Some of these extensions may not produce evenly
dispersed graph frequencies, as shown in \cite{shafipour2018directed}, and others
may produce degenerate eigenspaces.  A more general approach
is to use the Jordan decompostion \cite{sandryhaila2014discrete}, but this is numerically unstable and can produce transforms that do not properly preserve the notion of DC signals nor the overall energy (i.e., is a non-unitary transformation). 

An alternative motivation for defining a set of transform basis vectors and a
corresponding notion of frequency for directed, nonnegative graphs was
developed in \cite{shafipour2018directed}.  
This generalized the notion of the
\ac{TV} of a signal defined on an undirected, nonnegative graph defined as
\begin{equation}
    \text{TV}(\bm{x}) = \bm{x}^\top
    \bm{L}\bm{x}=\sum_{i,j=1,j>i}^NA_{ij}(x_i-x_j)^2\,.
\end{equation}
The primary insight is that $\text{TV}(\bm{v}_i)=\lambda_i$ for eigenvectors of
$\bm{L}$ and that the right hand side of the expression is amenable to generalization for the case of nonnegative directed graphs.  To this end,
\cite{shafipour2018directed} introduces the concept of \ac{DV} which is defined 
as
\begin{equation}
    \text{DV}(\bm{x})=\sum_{i,j=1}^NA_{ij}[x_i-x_j]^2_+\,,
    \label{eq_DV}
\end{equation}
where $[x]_+=\max\{0,x\}$. Note that for undirected graphs
$\text{DV}(\bm{x})=\text{TV}(\bm{x})$.
The intuition behind this quantity is that if directed edges of a graph
$\mathcal{G}$ represent the directed flow of a signal from higher values to
lower ones, only net positive signal flows will contribute to \ac{DV}. 

Having motivated \ac{DV}, \cite{shafipour2018directed} used intuition from
\ac{TV} to define the graph frequency of a unit vector $\bm{u}$ as
$f\triangleq\text{DV}(\bm{u})$, and thus for an arbitrary orthogonal matrix
$\bm{U}$, a directed \ac{GFT} is defined as $\hat{\bm{x}}=\bm{U}^\top \bm{x}$, which is associated with a set of frequencies $f_k$ corresponding to each column $\bm{u}_k$ through $f_k=\text{DV}(\bm{u}_k)$.  The majority of \cite{shafipour2018directed} then focuses on optimization routines for finding $\bm{U}$ that result in frequency components $f_k$ that are evenly spread.


\section{Directed Variation for Indefinite and Complex Graphs}

\subsection{Indefinite Directed Variation}

Unlike the work of \cite{shafipour2018directed} which focused primarily on approaches for analyzing graph functions on directed graphs with $A_{ij}\geq0$, in this section we first focus on adapting the techniques of \cite{shafipour2018directed} to the case where $A_{ij}$ can be both positive and negative, and $A_{ij} \neq A_{ji}$.  In this case, the Laplacian can have both negative, positive, or complex eigenvalues and non-orthogonal eigenvectors.
This suggests that we need to adapt \ac{DV} to properly
account for the variation introduced by the negative components of $A_{ij}$.
%
The natural adaptation is to extend
\ac{DV} to \ac{IDV} via
\begin{equation}
    \text{IDV}(\bm{x})=\sum_{i,j=1}^N[A_{ij}]_+[x_i-x_j]_+^2+[A_{ij}]_-[x_i-x_j]_-^2
\end{equation}
where $[x]_-= -\min\{0,x\}$.  This is equivalent to \ac{DV} for positive (or
negative) directed graphs, and thus is equivalent to \ac{TV} for undirected
graphs.

\subsubsection{Complex DV}

Note that the definition of \ac{IDV} is readily extendable to a notion of directed
variation for analysis when the graph signal or the directed adjacency matrix has
complex values.  Example use cases include multi-agent systems
\cite{lin2014distributed}, infrastructure networks
\cite{sanchez2013ict}, and neural networks \cite{frady2019robust}. Let
$\Re(\cdot)$ and $\Im(\cdot)$ denote the real and imaginary parts of the
argument, respectively, and for complex $\bm{A}$ and $\bm{x}$ let
\begin{equation}\begin{aligned}
    \tilde{\bm{y}}&=\begin{bmatrix}\Re(\bm{A}) & -\Im(\bm{A})\\ \Im(\bm{A}) &
        \Re(\bm{A}) \end{bmatrix}\begin{bmatrix}
    \Re(\bm{x})\\\Im(\bm{x})\end{bmatrix}&\triangleq \tilde{\bm{A}}\tilde{\bm{x}}
\end{aligned}\end{equation}
and then
\begin{equation}
    \bm{A}\bm{x}=\tilde{\bm{y}}_{1:N}+i\tilde{\bm{y}}_{N+1:2N}\,,
\end{equation}
where the subscripts indicate the first and second $N$ dimensions of $\bm{y}$,
respectively.  This equivalence between complex and real matrix algebras offers
an immediate generalization of \ac{DV} to the case of complex graphs
and graph signals by defining the \ac{CDV} of a complex signal as the \ac{IDV}
of the associated real valued adjacency matrix and graph signal. Formally,
given a complex, directed adjacency matrix $\bm{A}$ and graph signal $\bm{x}$,
define
%
\begin{align}
    &\text{CDV}(\bm{x})=\sum_{i,j=1}^{2N}[\tilde{A}_{ij}]_+[\tilde{x}_i-\tilde{x}_j]_+^2
    +[\tilde{A}_{ij}]_-[\tilde{x}_i-\tilde{x}_j]_-^2\notag\\
    &=\sum_{i,j=1}^N
    [\Re(A_{ij})]_+[\Re(x_i-x_j)]_+^2+[\Re(A_{ij})]_-[\Re(x_i-x_j)]_-^2\notag\\
    &+[\Re(A_{ij})]_+[\Im(x_i-x_j)]_+^2+[\Re(A_{ij})]_-[\Im(x_i-x_j)]_-^2\notag\\
    &+
    [\Im(A_{ij})]_-[\Re(x_i)-\Im(x_j)]_+^2\notag\\&+[\Im(A_{ij})]_+[\Re(x_i)-\Im(x_j)]_-^2\notag\\
    &+
    [\Im(A_{ij})]_+[\Im(x_i)-\Re(x_j)]_+^2\notag\\&+[\Im(A_{ij})]_-[\Im(x_i)-\Re(x_j)]_-^2
\end{align}

\subsection{Minimizing Spectral Dispersion}

Two methods for optimizing the spread of \ac{DV} were introduced in
\cite{shafipour2018directed}. The first, called the \textit{feasible} method,  used gradient descent on Stiefel
manifolds \cite{edelman1998geometry} which can be computationally prohibitive
due to the matrix decompositions for large graphs, as well nonconvexity of the
overall optimization problem requiring multiple initial conditions for the
gradient descent.  As an alternative, \cite{shafipour2018directed} also
introduced a \textit{greedy} heuristic that exploits submodularity which is both highly
efficient and uses basis vectors of a related undirected graph (this latter
fact is desirable as the resulting graph transform will be in a basis that is
in some sense ``natural''),  Next, we will discuss modifications to these
two approaches that are necessary to use \ac{IDV} and \ac{CDV} in place of \ac{DV}.

\subsubsection{Feasible Gradient Descent Approach}
The feasible gradient descent approach for minimizing spectral dispersion using \ac{IDV} proceeds
almost identically to what is described in \cite{shafipour2018directed}.  The
main differences are to replace \ac{DV} with \ac{IDV} in the objective function
computations.  This also changes the gradient computations, with the
only substantial change occurring to the single vector gradient $\bar{g}_i$
defined in \cite[Eq.~(15)]{shafipour2018directed}.  From the linearity of the
derivative, in the case of \ac{IDV} this becomes 
\begin{equation}\begin{aligned}
    \bar{g}_i =  &2\biggl([\bm{A}^\top_{\cdot
    i}]_+[\bm{u}-u_i\bm{1}_N]_+-[\bm{A}_{i\cdot}]_+[u_i\bm{1}_N-\bm{u}]_+\\
    &-[\bm{A}^\top_{\cdot
    i}]_-[\bm{u}-u_i\bm{1}_N]_-+[\bm{A}_{i\cdot}]_-[u_i\bm{1}_N-\bm{u}]_-\biggr)
\end{aligned}\end{equation}
to use the notation of \cite{shafipour2018directed}.

An extension of the gradient descent
approach for \ac{CDV} is also straight-forward.  We can compute the single-vector complex
gradient of the now complex vector by using the same complex-to-real transform used to derive \ac{CDV} from \ac{IDV}.
For a complex adjacency matrix $\bm{A}$ and vector $\bm{u}$ (and transformed
real-valued $\tilde{\bm{A}}$ and $\tilde{\bm{u}}$) we can define an associated
real-valued gradient $\tilde{\bar{\bm{g}}}$ by 
\begin{equation}\begin{aligned}
    \tilde{\bar{g}}_i =  &2\biggl([\tilde{\bm{A}}^\top_{\cdot
    i}]_+[\tilde{\bm{u}}-\tilde{u}_i\bm{1}_{2N}]_+-[\tilde{\bm{A}}_{i\cdot}]_+[\tilde{u}_i\bm{1}_{2N}-\tilde{\bm{u}}]_+\\
    &-[\tilde{\bm{A}}^\top_{\cdot
    i}]_-[\tilde{\bm{u}}-\tilde{u}_i\bm{1}_{2N}]_-+[\tilde{\bm{A}}_{i\cdot}]_-[u_i\bm{1}_{2N}-\tilde{\bm{u}}]_-\biggr)
\end{aligned}\end{equation}
and then the complex gradient is
$\bar{g}=\tilde{\bar{g}}_{1:N}+i\tilde{\bar{g}}_{N+1:2N}$. The remaining
changes to gradient descent approach are handled by taking the appropriate
inner product on the complex space (i.e., using conjugate transpose instead of
transpose), as outlined in the original reference \cite[Sec.~4.2]{wen2013feasible}
to the feasible method used by \cite{shafipour2018directed}.

\subsubsection{Greedy Heuristic Approach}
Due to the computational complexity and non-convexity of the gradient descent
approach, \cite{shafipour2018directed} introduced a greedy heuristic that uses
the Laplacian of a corresponding undirected graph.
Specifically, from a positive directed graph $\mathcal{G}=(\mathcal{V},\bm{A})$
they define the \textit{underlying undirected graph}
$\mathcal{G}^u=(\mathcal{V}, \bm{A}^u)$ where
$A_{ij}^{u}=\max\{A_{ij},A_{ji}\}$. Since $\mathcal{G}^u$ is symmetric, the
corresponding Laplacian $\bm{L}^u$ will have orthogonal eigenvectors and can be used for a directed \ac{GFT}. An additional motivation
for considering $\mathcal{G}^u$ is that \ac{DV}$(\bm{x})$ (with respect to
$\mathcal{G}$) is bounded by $\lambda_{max}^u$, the maximum eigenvalue of
$\bm{L}^u$.  Noting that for directed graphs
$\text{\ac{DV}}(\bm{x})\neq\text{\ac{DV}}(-\bm{x})$, the authors use a greedy
search over the collection $\{\pm{\bm{v}_i}\}$ for $\bm{v}_i$ eigenvectors of
$\bm{L}^u$ that chooses only one from $\pm \bm{v}_i$ for each $i$.  This relies on
a proof of submodularity of the spread of \ac{DV} derived using matroid
theory.

In the case of \ac{IDV}, we use the underlying graph with adjacency matrix $\bm{A}^{|u|}=\max\{|A_{ij}|,|A_{ji}|\}$ and the eigenvectors of the corresponding Laplacian $\bm{L}^{|u|}$. The matroid conditions still hold, as the basis elements of $\bm{L}^{|u|}$ are orthonomal.  Thus, the spread of
\ac{IDV} will also be submodular, and the greedy optimization of \ac{IDV}
spread can proceed as described in \cite{shafipour2018directed}, albeit with
the subsitution of \ac{IDV} in place of \ac{DV}.  The presence of complex
weights presents an additional wrinkle in the greedy optimization approach.  In
the complex case, we should instead consider the effect of an arbitrary
unit-norm complex scalar $e^{i\theta}$ on the basis vectors, as opposed to
simply $\pm1$.  This presents a scalar-argument continuous optimization problem
that must be computed for each basis vector at each time step of the greedy
approach.  Furthermore, this optimization is not convex so will require several
initial conditions to find the global minimum in a gradient descent approach.
As an alternative, we propose instead to compute initially the \ac{CDV} of each
basis vector multiplied by $e^{i\theta_k}$ for $\theta_k$ evenly spaced on a
grid between $[0,2\pi)$ and then proceeding with the greedy algorithm, with the
selection made over rotation angles $\theta_k$ as opposed to only $\pm1$.

\section{Results}

\subsection{Simulation Results}
%
We performed the following experiment to assess the impact of using the appropriate notion of \ac{DV}. We considered ring lattice networks with $N=16$ nodes and degree 2, Erdos Renyi networks with $N=16$ nodes and probability of attachment $p=0.2$, and stochastic block networks with three communities and $N_c=8$ nodes per community.  All edges were directed with weights uniformly drawn from $\{\pm1,\pm i\}$.  For each of these graphs $\mathcal{G}$, we created an indefinite graph $\mathcal{G}_I$ by replacing the $\pm i$ edges with $\pm1$, and a positive graph $\mathcal{G}_P$ by setting all weights to $1$.  For 10,000 random instances of the above classes, we generated two random unit vectors in $\mathbb{R}^N$ and compared the resulting \ac{DV}s using $\mathcal{G}_P$ to the \ac{IDV}s using $\mathcal{G}_I$ and the \ac{CDV}s using $\mathcal{G}$.  Across all 60,000 comparisons we found that the relative orderings of the two vectors were different greater than 35\% of the time.  This highlights the importance of \ac{IDV} or \ac{CDV}, as the frequency interpretation of a given basis element or signal can change substantially.

\begin{figure}[h!]
    \centering
    \begin{tabular}{cc}
    \subfloat[]{\includegraphics[width=.45\columnwidth]{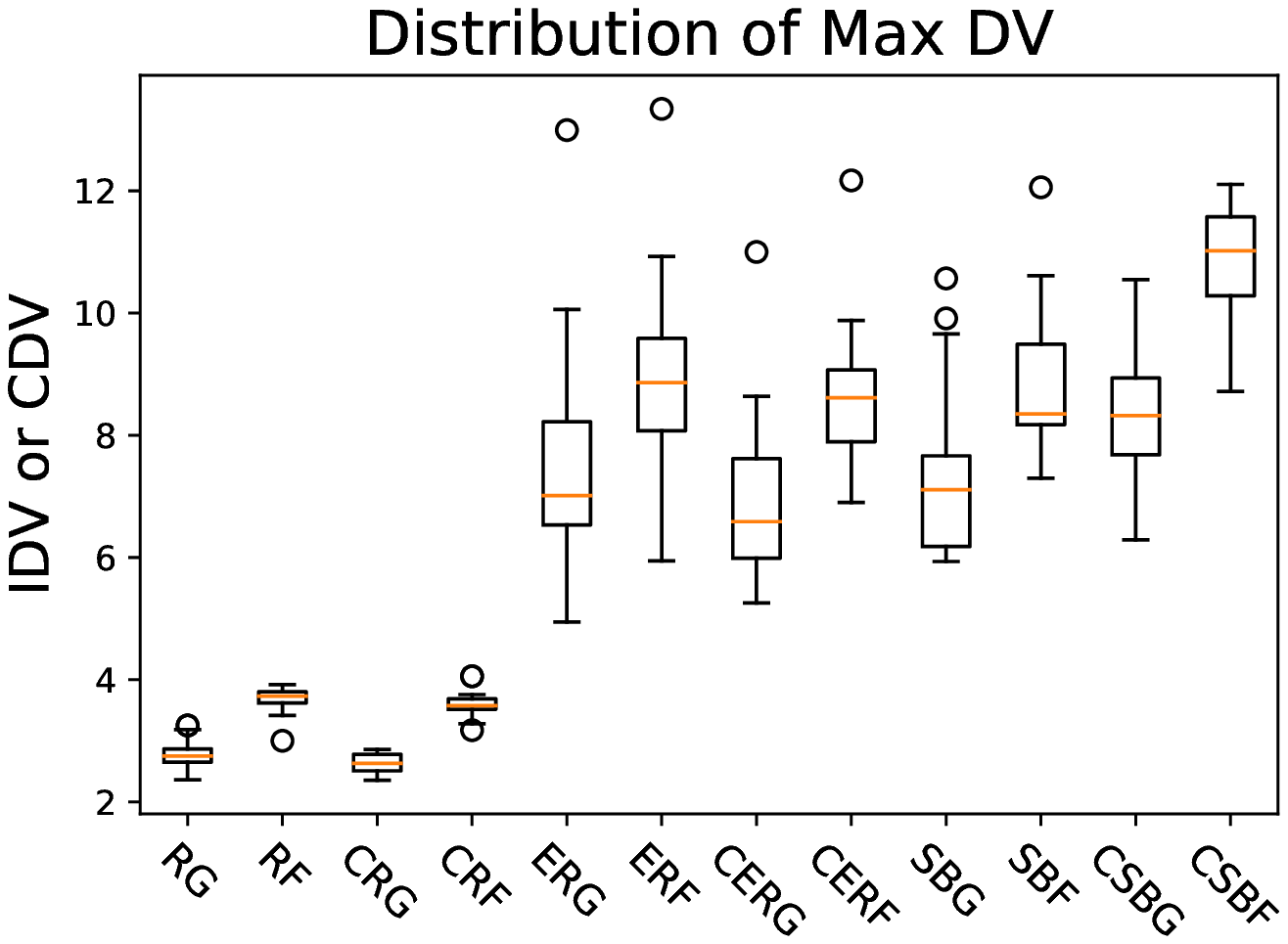}\vspace{1cm}\label{fig:sims:a}}
    \subfloat[]{\includegraphics[width=.45\columnwidth]{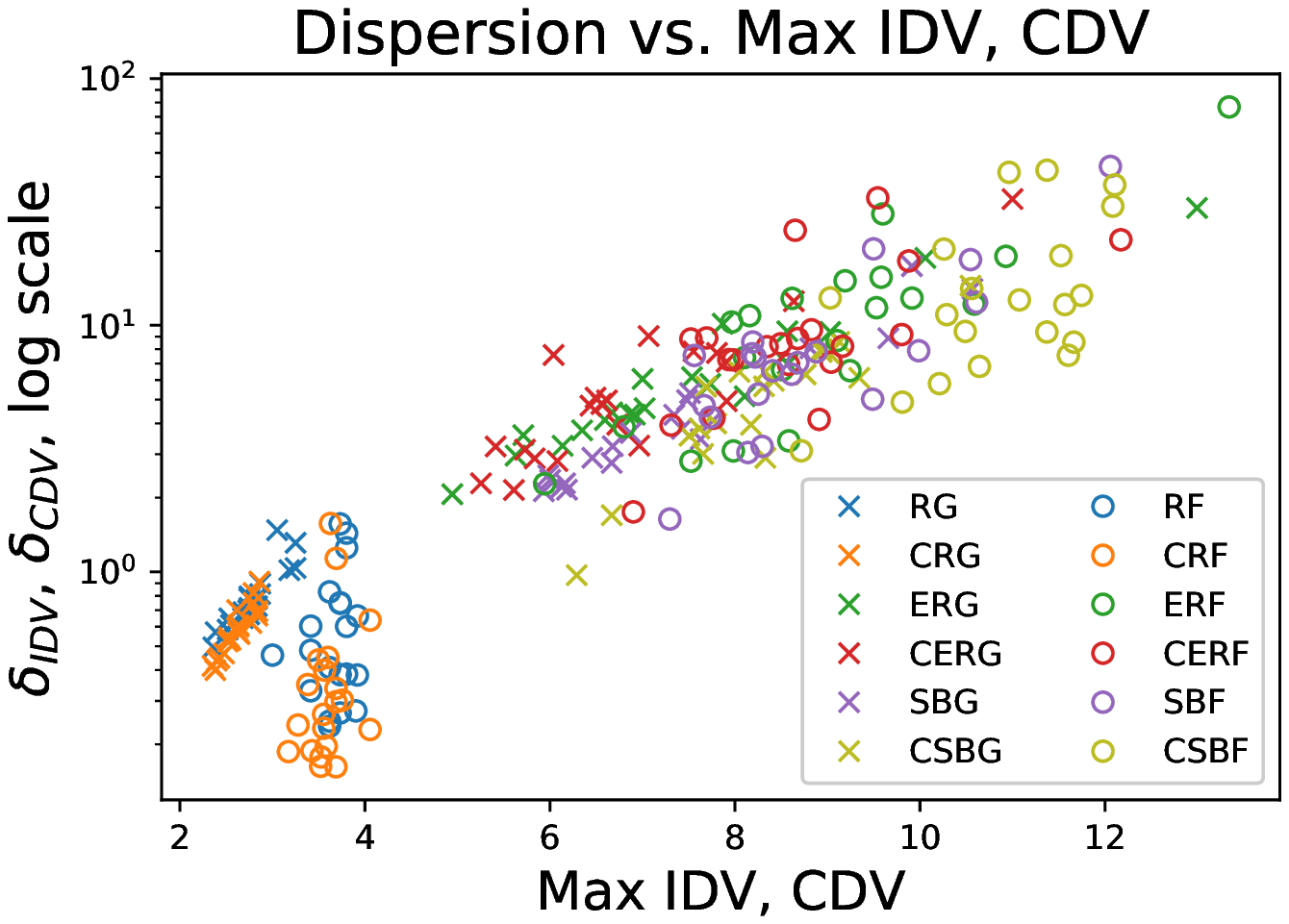}\label{fig:sims:b}}
    \end{tabular}
    \caption{\protect\subref{fig:sims:a} Distributions of maximum \ac{IDV} or \ac{CDV} basis elements. $R$, $ER$, and $SB$  denote the graph class, a prefix $C$ denotes complex, and a $G$ or $F$ suffix denotes greedy or feasible method. \protect\subref{fig:sims:b} Scatter plots of dispersion vs.\ maximum \ac{IDV}/\ac{CDV} to illustrate difficulty comparing dispersion between graphs and \ac{GFT}s.
    }
    \label{fig:sims}
\end{figure}

To further understand the proposed methods we compared them on a subset of $M=20$ for graph class above.  Fig.~\ref{fig:sims:a} shows box-plots of the maximum \ac{IDV} and \ac{CDV}, indicating that the feasible method increases \ac{DV} beyond the greedy method.  Furthermore, these show that distributions of particular classes of graphs can vary widely. Unlike maximum \ac{DV}, it is essentially impossible to directly compare dispersion, defined as $\delta_{IDV}(\bm{U}) = \sum_{i=1}^{N-1} [\text{IDV}(\bm{u}_{i+1}) - \text{IDV}(\bm{u}_i) ]^2$ (and similarly $\delta_{CDV}$) as dispersion is strongly correlated with maximum \ac{DV} (see Fig.~\ref{fig:sims:b}), meaning the feasible method also tends to produce an increase in dispersion, despite a qualitatively smoother distribution.  That said, each graph class does tend to produce relatively similar dispersions.

\subsection{Fruit Fly Protocerebral Bridge}
\begin{figure}[h!]
    \centering
    \begin{tabular}{cc}
    \subfloat[]{\includegraphics[width=.45\columnwidth]{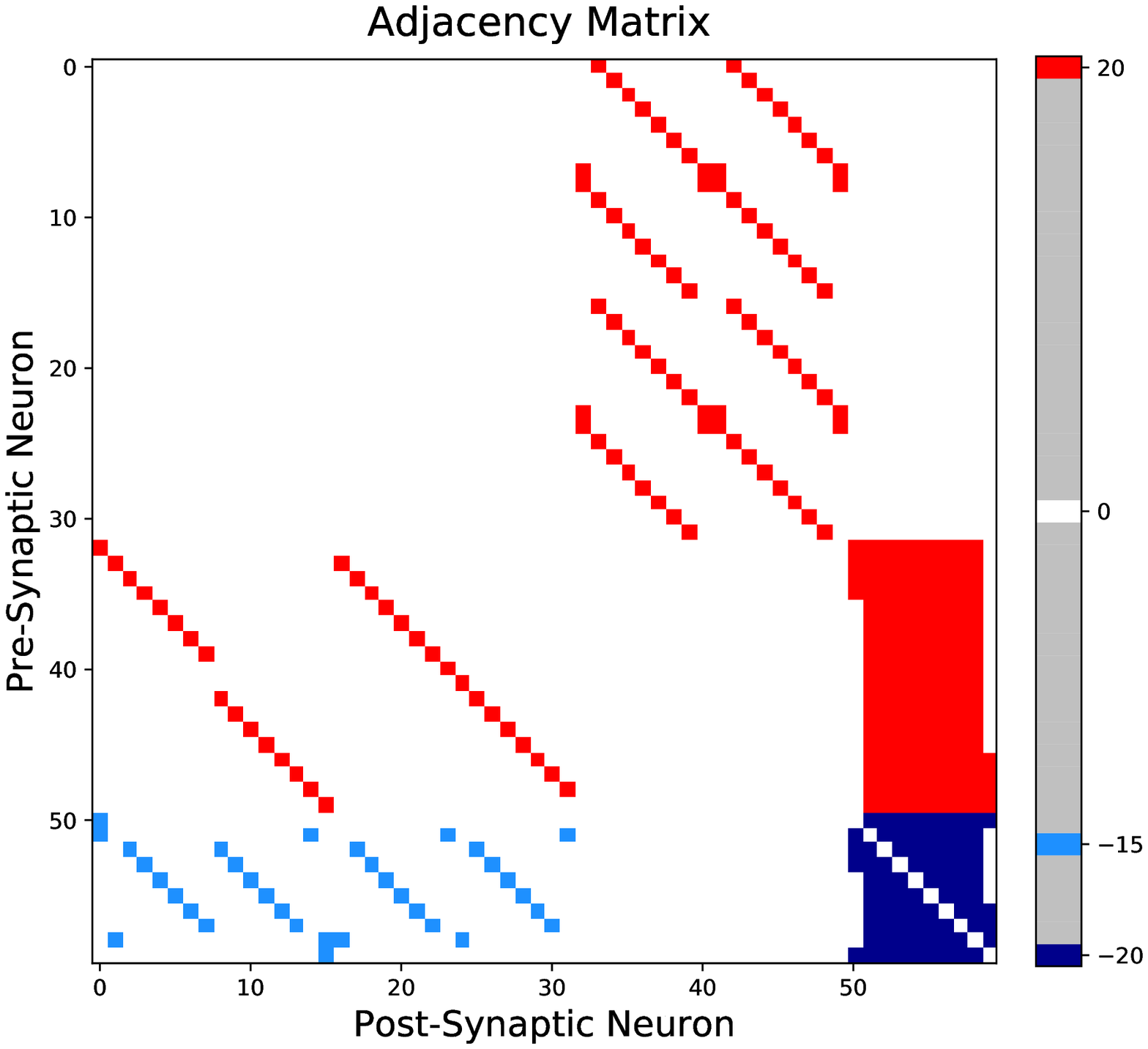}\vspace{1cm}\label{fig:fly_adj:a}}
    \subfloat[]{\includegraphics[width=.45\columnwidth]{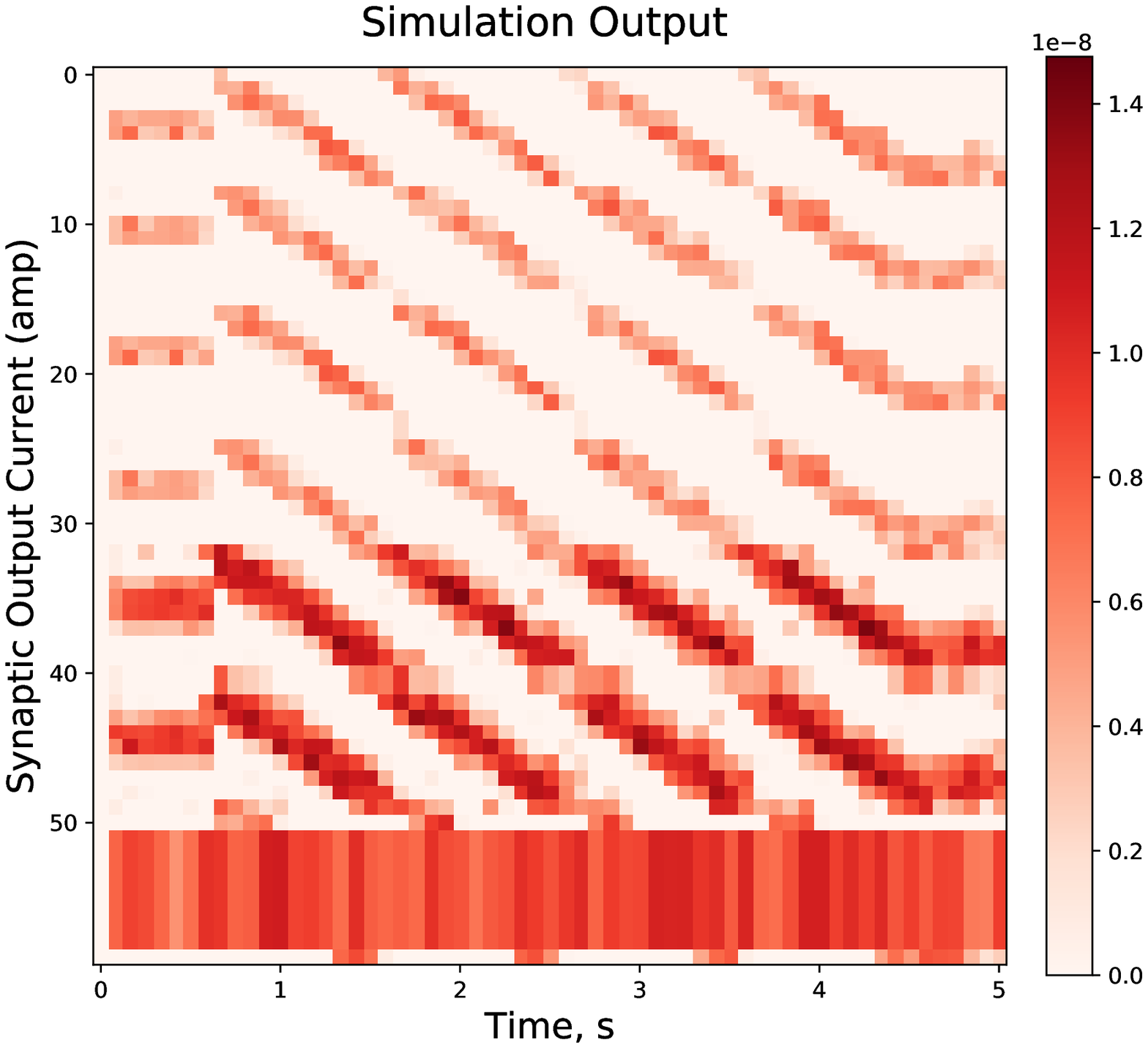}\label{fig:fly_adj:b}}
    \end{tabular}
    \caption{\protect\subref{fig:fly_adj:a} Adjacency matrix of the fruit-fly \ac{PB} model from \cite{kakaria2017ring}. Note the presence of both excitatory neurons (red)
    and inhibitory neurons (blue). \protect\subref{fig:fly_adj:b} Sample simulation under noisy fixed frequency rotation stimulus into the \ac{PB} from 0.5-4.5~s as described in \cite{kakaria2017ring}. For presentation purposes, every 833rd sample of the fixed sample time (0.1~ms) simulation is shown.
    }
    \label{fig:fly_adj}
\end{figure}

As biological neural networks were the primary motivator for developing
\ac{IDV}, in this example we consider a model of the fruit fly
(\textit{Drosophila melanogaster}) \ac{PB} from \cite{kakaria2017ring}. This sub-system of the fruit fly brain is thought to be responsible for maintaining heading direction in the navigation process of the fruit fly. The adjacency matrix and graph signals are provided from a recent study \cite{kakaria2017ring}. It includes three primary structures, the \ac{EB} (nodes 0-31), the excitatory portion of the \ac{PB} (nodes 32-49), and the inhbitory portion of the the \ac{PB} (nodes 50-59; see Fig.~\ref{fig:fly_adj} (a)) as well as a simulation that
produces ring attractor dynamics that we use to define the graph signals in our analysis.  The adjacency matrix in Fig.~\ref{fig:fly_adj} is
highly asymmetric, as synaptic signals only flow in one direction, and follows Dale's Law (i.e., each neuron affects others with only positive or
negative weights).  An example simulation is shown in Fig.~\ref{fig:fly_adj:b}, where the network is initially (0.5~s) subjected to a background noisy spiking current stimulus into the \ac{PB}, followed by 4~s of background stimulus plus feed-forward stimulus representing constant, periodic rotation of the fly's heading, followed by another 0.5~s of background stimulus.
%
%
Other than the repeating stimulus, we use the default parameters from \cite{kakaria2017ring}.

\begin{figure}[ht!]
    \centering
    \begin{tabular}{cc}
    \subfloat[]{\includegraphics[width=.45\columnwidth]{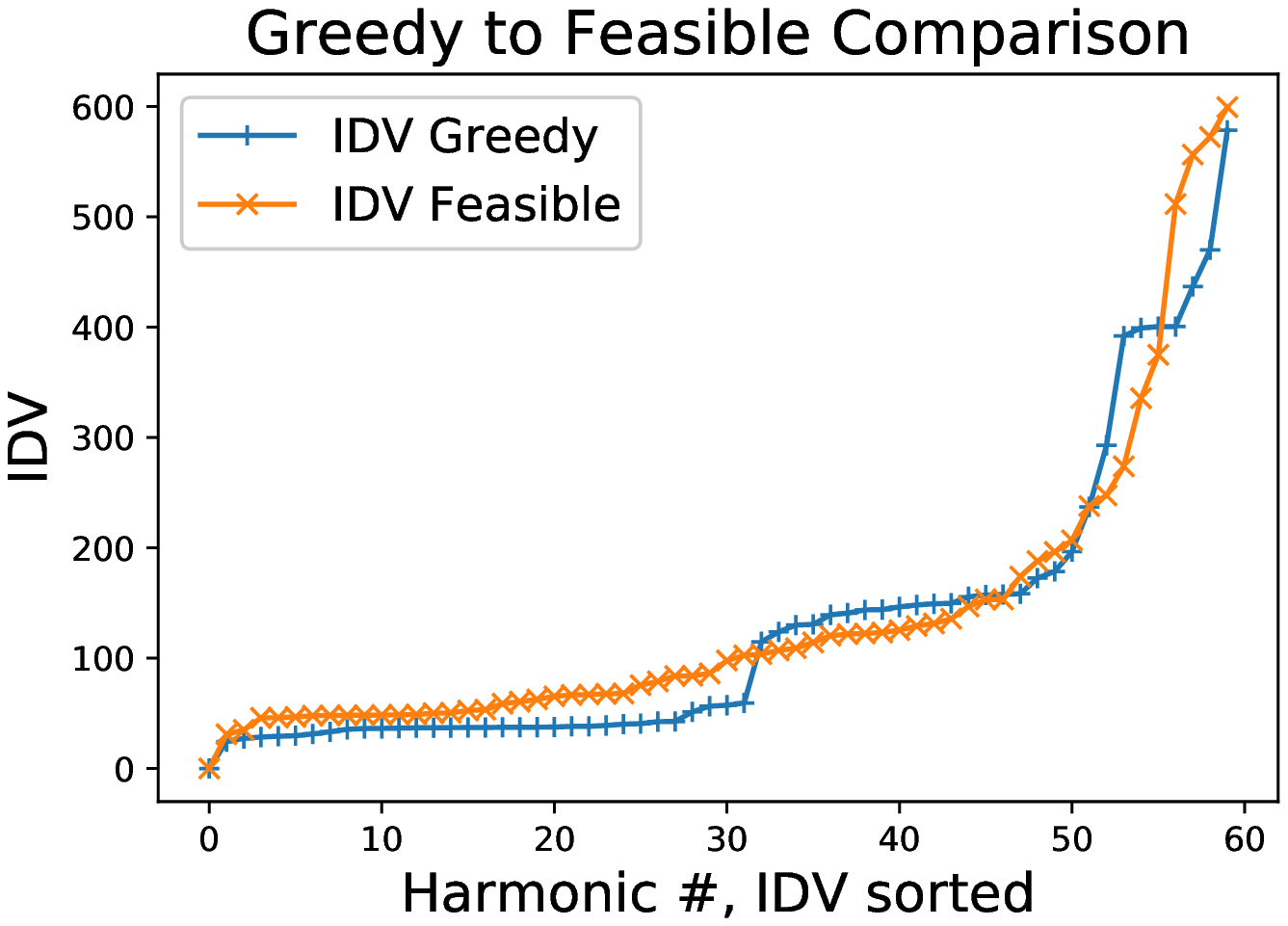}\label{fig:idv_dv:a}}
    \subfloat[]{\includegraphics[width=.45\columnwidth]{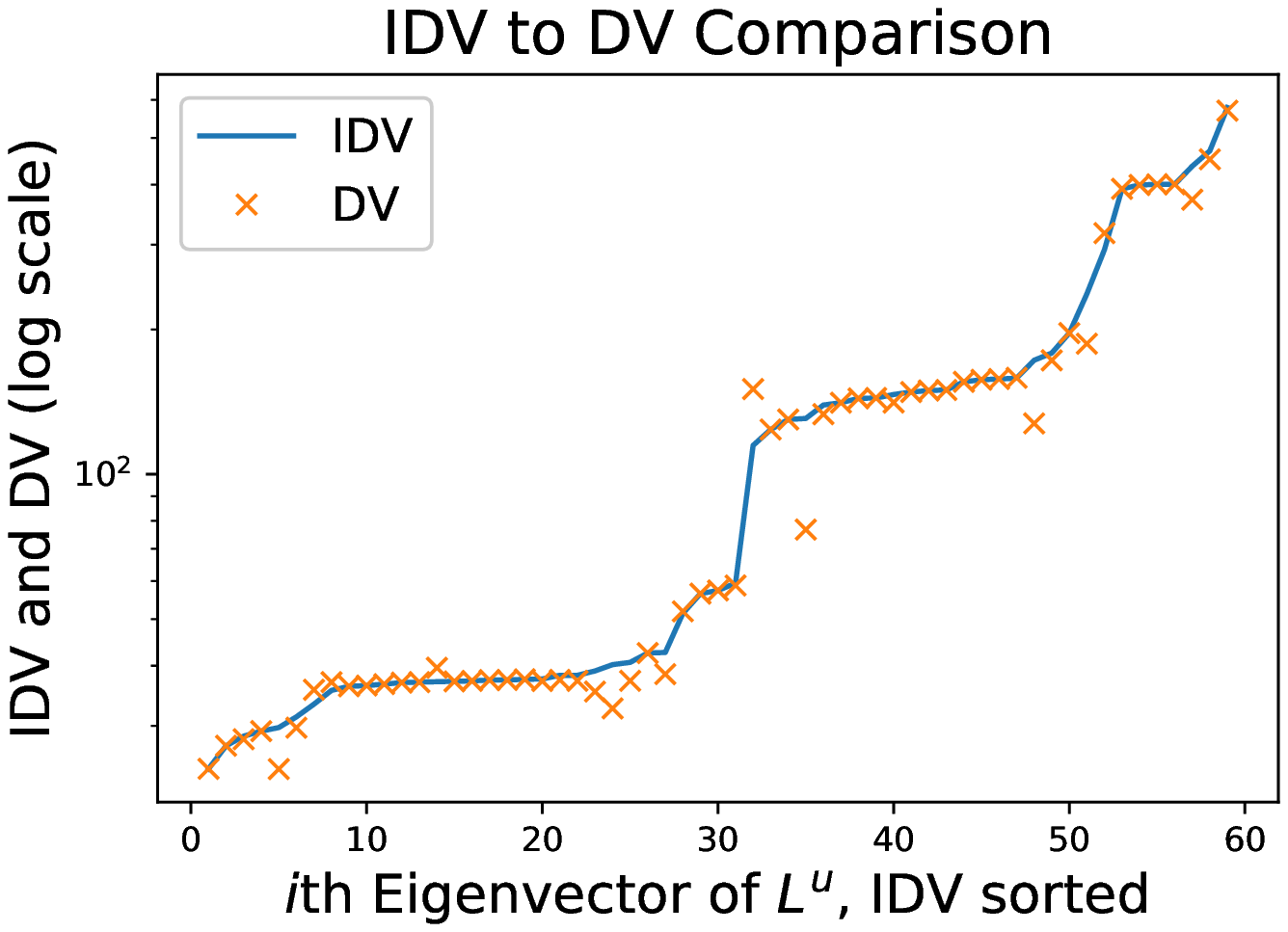}\label{fig:idv_dv:b}}&
    \end{tabular}
    \caption{\protect\subref{fig:idv_dv:a} Comparison of \ac{IDV} values resulting from
    greedy and feasible approaches.  Note higher maximum and qualitatively smoother (more 
    even) distribution for the feasible method. \protect\subref{fig:idv_dv:b} Comparison 
    of \ac{IDV} and \ac{DV} values from greedy methods, sorted by \ac{IDV} value of 
    corresponding underlying Laplacian eigenvector. 
    }
    \label{fig:idv_dv}
    
\end{figure}

Fig.~\ref{fig:idv_dv:a} shows the resulting \ac{IDV}s from the greedy and feasible methods. The feasible method results in more even spacing in \ac{IDV}s compared to the greedy. 
Comparisons between \ac{IDV} and \ac{DV} from greedy methods are shown in Fig.~\ref{fig:idv_dv:b}, showing considerable differences in the relative ordering of the identical harmonics despite identical basis elements.  This indicate substantially different frequency interpretation under the two measures, and thus the inclusion of negative weights in the graph is meaningful. It is more difficult to directly compare the feasible methods, as the graph harmonics produced are  quite different. One clear point of comparison, however, is that between the different ``maximum frequency'' harmonics produced by the first stage of the feasible method. Table~\ref{tab:ff_compare} shows that while the greedy methods agree on the maximum frequency harmonic, the feasible methods produce quite different results. Again, this further illustrates the importance of including both directivity and weight into the \ac{GFT}.

\begin{table}[ht!]
    \centering
    \caption{ Comparison between greedy and gradient descent approaches}
    \begin{tabular}{|c|c|c||c|c|}\hline
        & Feas. IDV & Greedy IDV & Feas. DV & Greedy DV\\\hline
        \textbf{Max IDV}& 599.43 & 578.48 & 576.24 &578.48\\\hline
        \textbf{Max DV} &559.51 &569.86 & 570.39 &569.86\\\hline
        \textbf{$\delta_{IDV}$, $\delta_{CDV}$}& 31286.5 & 33420.8 & 28780.3 &38559.9\\\hline

    \end{tabular}

    %

    \label{tab:ff_compare}

\end{table}

\begin{figure}[ht!]
    \centering
    \begin{tabular}{cc}
    \subfloat[]{\includegraphics[width=.45\columnwidth]{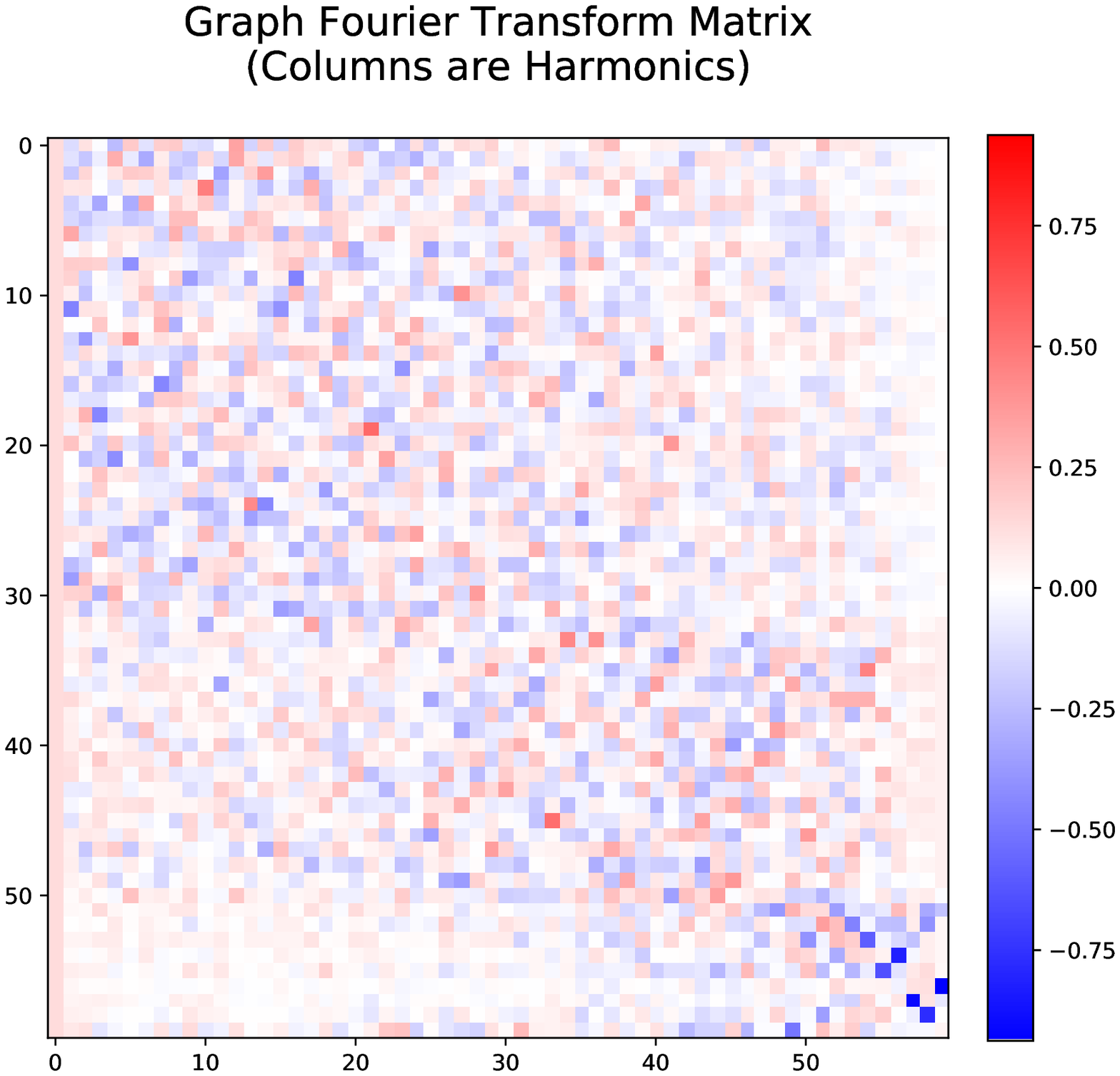}\label{fig:fly_xform:a}}
    \subfloat[]{\includegraphics[width=.45\columnwidth]{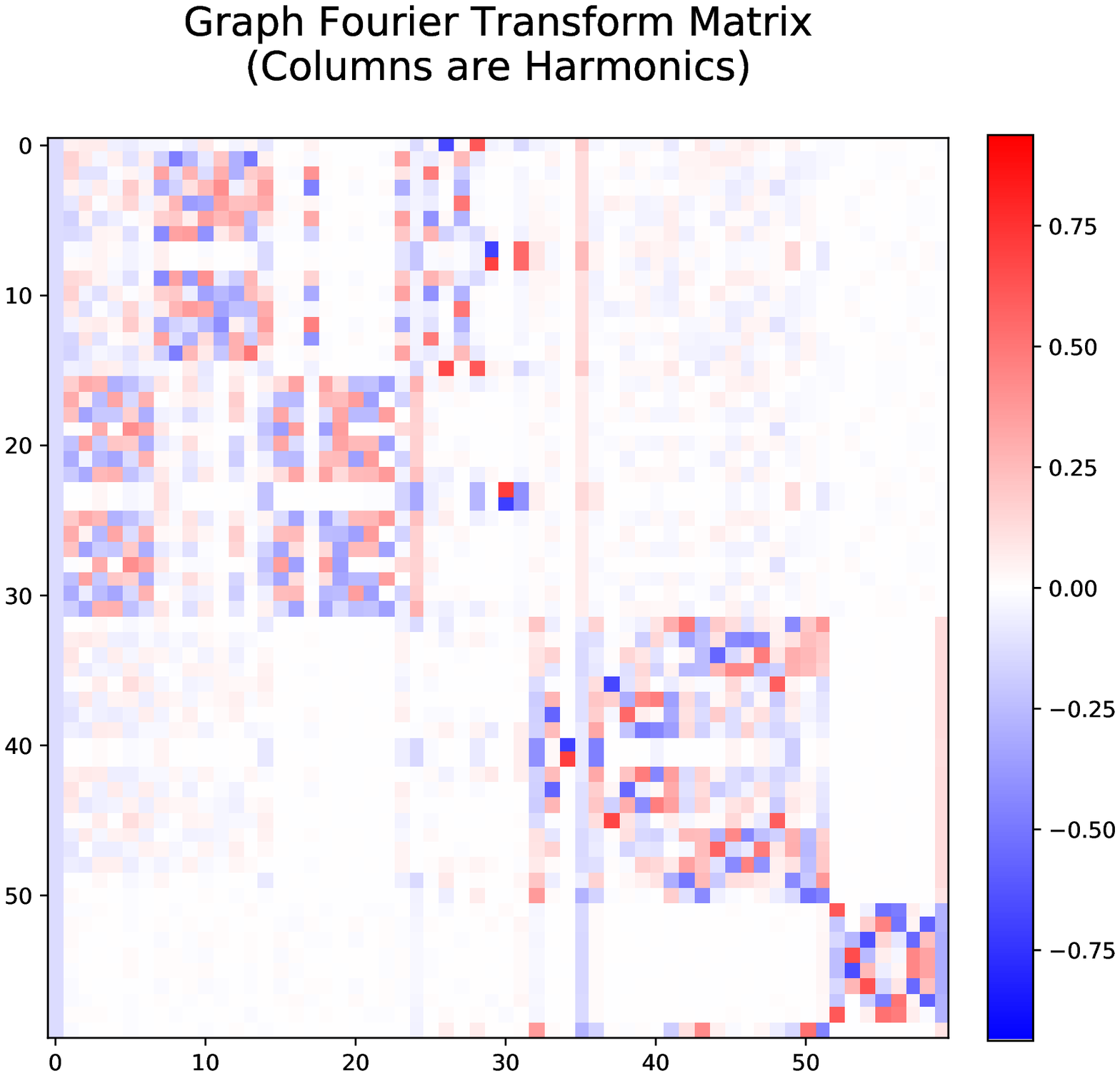}\label{fig:fly_xform:b}}&
    \end{tabular}
    \caption{\ac{IDGFT} matrices for adjacency matrix in Fig.~\ref{fig:fly_adj}.  \protect\subref{fig:fly_xform:a} Feasible gradient descent approach. \protect\subref{fig:fly_xform:b} Greedy submodular approach.  While the feasible method produces both a higher maximum \ac{IDV} and lower overall \ac{IDV} dispersion, the greedy approach ``captures'' more of the network structure.}
    \label{fig:fly_xform}
    
\end{figure}

IDGFT matrices generated using the feasible and greedy \ac{IDV} methods are shown in Fig.~\ref{fig:fly_xform:a} and Fig.~\ref{fig:fly_xform:b}, respectively, with eigenvectors sorted by \ac{IDV}.  Despite the perceived benefits of the feasible method for this network, we find that the greedy transform more intuitively captures the structure of the network leading to more interpretable graph Fourier analysis, whereas the feasible method spreads the energy more evenly across the different harmonics. Furthermore, the ``discontinuties'' in the greedy \ac{IDV} values (Fig.~\ref{fig:idv_dv:a}) actually correspond to important functional blocks within the highly structured network considered here.

\begin{figure}[h!]
    \centering
    \includegraphics[width=.8\columnwidth]{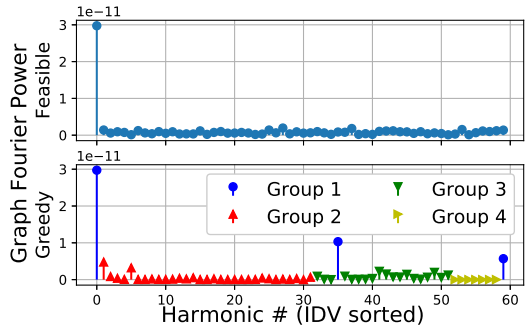}
    \caption{Power spectra of graph signal from Fig.~\ref{fig:fly_adj:b}. Top: Feasible IDV-based transform. Bottom: Greedy IDV-based transform, colored by frequency groupings. Group 1 corresponds to harmonics that capture coarse activity levels between the \ac{EB} (Group 2), excitatory-\ac{PB} and boundary inhibitory neurons (Group 3) and interior inhibitory neurons (Group 4).}
    \label{fig:graph_power}

\end{figure}

The total \ac{GFT} power of the graph signal (over time) in Fig.~\ref{fig:fly_adj:b} using the two transforms in Fig.~\ref{fig:fly_xform} is shown in Fig.~\ref{fig:graph_power}.  Using visual inspection in conjunction with the \ac{IDV} distribution, we have divided the greedy harmonics into four groups as shown in Fig.~\ref{fig:graph_power}~(bottom). The first grouping includes basis elements 0, 35, and 59 which coarsely measures signal in the entire network, in the \ac{EB} (nodes 0-31) vs.\ the \ac{PB} (32-59), and the interior inhibitory neurons (51-58) vs.\ the remainder of the \ac{PB}. These three elements coarse-grain the signal power into these functional blocks. The second grouping covers spectral content primarily within the the \ac{EB}.
Curiously, the third grouping includes not only the excitatory neurons in the \ac{PB}, but also the first and last inhibitory neurons, presumably due to the fact that they have a slightly different neighborhood structure from the other inhibitory neurons.  This also leads to a fundamentally different graph signal for those boundary neurons, as evident in Fig.~\ref{fig:fly_xform:b}.  The final group consists of the interior inhibitory neurons which appear to add little additional spectral content to the graph beyond their average (captured in the first group).


%
%

\begin{figure}[h!]
\centering
\includegraphics[width=.8\columnwidth]{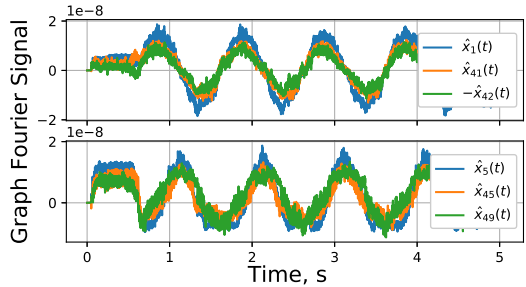}
\caption{Fluctuations of projections of graph signals into certain graph harmonics that exhibit periodic behavior consistent with the input rotation rate.  Note the bottom signals are 90$^\circ$ out of phase with those in the top.  This signals, correspond to peaks in Group 2 and 3 in Fig.~\ref{fig:graph_power} and comprise $\approx$20\% of the total graph signal power.}

\label{fig:time_harmonics}

\end{figure}

An interesting facet revealed through this graph Fourier analysis is that the two dominant components in the ellipsoid body, $\hat{x}_1(t)$ and $\hat{x}_5(t)$, result in periodic oscillations (at the input rotation period) that are 90$^\circ$ out of phase (see Fig.~\ref{fig:time_harmonics}).  As the inputs to the system are in the \ac{PB}, these dynamics must be driven from there, and indeed we see that the top four components have the same periodic structure ($\hat{x}_{41}$, $\hat{x}_{42}$ $\hat{x}_{45}$ and $\hat{x}_{49}$).  Since the \ac{EB} then feeds back into the \ac{PB} it would seem to indicate that the \ac{EB} effectively aggregates and stabilizes several Fourier components and feeds them back in the reciprocal relationship between the \ac{EB} and the \ac{PB}.

%


\section{Conclusions and Future Work}
In summary, we have extended the techniques in \cite{shafipour2018directed} to develop a \ac{GFT} that can be used for indefinite- and complex-weighted graphs.  We showed that much of the intuition and rationale behind the original \ac{DV} motivation applies to the transforms presented here.  Furthermore, as using only the absolute value of the weights can have a substantial impact on the frequency interpretation of graph harmonics and the resulting analysis, this indicates the need to consider both the sign or phase of the weights in addition to the directionality.  Finally, we applied these tools to a simulated biological neural network and were able to apply \ac{GFT} analysis to derive insight into its dynamics. 
In future work, we will explore other algorithms and heuristics for the optimization of IDV and CDV, and apply IDGFT and CDGFT to other applications.

\section*{Acknowledgement}
We thank Dr. Grace Hwang for pointing us to the fly simulation and for helpful discussions.
%

\bibliographystyle{IEEEtran}
\bibliography{references}

\end{document}